\pdfoutput=1
 
 \RequirePackage{fix-cm}
\documentclass[10pt,conference]{IEEEtran}

\usepackage[fontsize=9.6]{fontsize}
\IEEEoverridecommandlockouts
\usepackage[nolist]{acronym}
\usepackage{cite}
\usepackage{hyperref}
\usepackage{amsmath,amssymb,amsfonts}
\usepackage{cleveref}

\usepackage{graphicx}
\usepackage{textcomp}
\usepackage{xcolor}
\usepackage{pbalance}
\usepackage{pgfplots}
\usepackage{tikz-dsp}
\usepackage{caption}
\usepackage{subcaption}
\usepackage{multirow}
\usepackage{mathabx}
\usepackage{algorithm}
\usepackage{algpseudocode}
\usepackage{enumerate}

\algblock{Input}{EndInput}
\algnotext{EndInput}
\algblock{Output}{EndOutput}
\algnotext{EndOutput}
\newcommand{\Desc}[2]{\State \makebox[2em][l]{#1}#2}

\usepgfplotslibrary{groupplots,dateplot}
\usetikzlibrary{patterns,
				shapes.arrows,
                shapes.multipart,
				arrows,
				backgrounds,
				fit,
				positioning,
				calc}
\pgfplotsset{compat=newest}

\def\BibTeX{{\rm B\kern-.05em{\sc i\kern-.025em b}\kern-.08em
    T\kern-.1667em\lower.7ex\hbox{E}\kern-.125emX}}
\let\oldtt\texttt
\renewcommand{\texttt}[1]{\small\oldtt{#1}}

\renewcommand{\verb}[1]{\small\oldverb#1}

\newlength{\footnoterulewidth} 
\newlength{\footnoteruleheight} 
\renewcommand{\footnoterule}{   
    \kern -3pt   \hrule width \footnoterulewidth height \footnoteruleheight   \kern \dimexpr 3pt - \footnoteruleheight \relax 
}

\Crefformat{figure}{#2Fig.~#1#3}
\Crefmultiformat{figure}{Figs.~#2#1#3}{ and~#2#1#3}{, #2#1#3}{ and~#2#1#3}

\newcommand{\sizex}{7}
\newcommand{\sizey}{7}
\newcommand{\sizexs}{4} % size for source image
\newcommand{\sizeys}{4}

\newcommand{\globalscale}{0.14000000}

\begin{document}

\begin{acronym}
    \acro{db}[dB]{decibel}
    \acro{EDSR}[EDSR]{Enhanced Deep
Residual Networks for single Image Super-Resolution}
    \acro{GIRRE}[GIRRE]{Guided \ac{IR} Resolution Enhancement}
    \acro{HR}[HR]{high-resolution}
    \acro{IR}[IR]{infrared}
    \acro{LR}[LR]{low-resolution}
    \acro{MISR}[MISR]{multi image super resolution}
    \acro{NIR}[NIR]{near infrared}
    \acro{PSNR}[PSNR]{peak signal-to-noise ratio}
    \acro{SISR}[SISR]{single image super resolution}
    \acro{SSIM}[SSIM]{structural similarity index measure}
    \acro{SSR}[SSR]{stereo super resolution}
    \acro{SWIR}[SWIR]{short-wave infrared}
    \acro{VDSR}[VDSR]{Accurate Image Super-Resolution using Very Deep
    Convolutional Networks}
    \acro{VDSR_NIR}[VDSRNIR]{VDSR trained on NIR images}
    \acro{VSR}[VSR]{video super resolution}
\end{acronym}

\title{RGB-Guided Resolution Enhancement of IR Images}

\author{
    \IEEEauthorblockN{
        Marcel Trammer$^{1,2}$, Nils Genser$^1$, and Jürgen Seiler$^2$
    }

    \IEEEauthorblockA{
        \\
        $^1$ AI \& Signal Processing Lab, Sielaff GmbH \& Co. KG,
        Herrieden, Germany \\
        $^2$ Multimedia Communications and Signal Processing, \\
        Friedrich-Alexander-Universität Erlangen-Nürnberg (FAU), Erlangen, Germany\\
        \\
        \texttt{\{m.trammer, n.genser\} @sielaff.de},
        \texttt{juergen.seiler @fau.de}
    }
}

\maketitle

\begin{abstract}
    This paper introduces a novel method for RGB-Guided Resolution
    Enhancement of infrared (IR) images
    called Guided IR Resolution Enhancement~(GIRRE).
    In the area of single image super resolution~(SISR) there exists a
    wide variety of algorithms like interpolation methods or neural
    networks to improve the spatial resolution of images.\ In
    contrast to SISR, even more information can be gathered on the recorded
    scene when using multiple cameras.
    In our
    setup, we are dealing with multi image super resolution, especially with
    stereo super resolution.\ We consider a color camera and an IR
    camera.
    Current IR sensors have a very low
    resolution compared to color sensors so that recent color sensors
    take up 100 times more pixels than IR sensors.
    To this end, GIRRE increases the spatial resolution of the
    low-resolution IR image.
    After that, the upscaled image is filtered with the aid of the
    high-resolution color image.
    We show that our method achieves an average PSNR gain of 1.2\,dB and at
    best up to 1.8\,dB compared to state-of-the-art methods, which is visually noticeable.
\end{abstract}


\section{Introduction}\label{sec:intro}
In distributed camera systems consisting of color cameras and
\ac{IR} cameras, the
spatial resolution typically differs by orders of magnitude.
Latest \ac{IR} sensors like Sony IMX 990 achieve a resolution of 1.3 MP~\cite{on:IMX990}.\
In contrast, modern CMOS sensors like Sony IMX 411AQR reach a resolution
of 151.4~MP~\cite{on:IMX411}.\ This
means, that modern color cameras can take up to 100 times as many
pixels as
\ac{IR} cameras.\ Because of the higher number of pixels recorded, the color
cameras record more spatial information about the scene than \ac{IR} cameras.\
By using color information as additional source, we propose to increase
the spatial resolution of \ac{IR} cameras.
Therefore, our novel \ac{GIRRE} algorithm first increases the spatial
resolution of the low-resolution \ac{IR} image.
Then, this newly generated image is filtered with the help of the
high-resolution color image to restore missing high-frequent information in
the \ac{IR} image.
Up to now, we assume our setup consisting of
one color camera and one \ac{IR} camera.\ Furthermore, we assume that the
images are already well registered, for example by methods
as proposed in~\cite{art:CAMSI,art:deep_material_aware}.

In order to understand what possibilities we have to increase the
spatial resolution, we will give an overview of the
state of the art in \Cref{sec:state}.\ In
\Cref{sec:prop_meth}, we introduce our novel method.
To demonstrate the improvement against the \mbox{state-of-the-art} methods, we
evaluate our method
with respect to different databases, multiple scale factors, and latest
\mbox{state-of-the-art} algorithms in~\Cref{sec:evaluation}.

\section{State of the art}\label{sec:state}
Today's smartphones are usually equipped with several
cameras~\cite{pat:camera_module_including,pat:dual_camera_module}.
When using multiple cameras, it is obvious to equip them with different
modalities to achieve more information of the recorded scene.
For example, a combination of a color camera and an \ac{IR} camera is often
used.
One of the most common examples is the Apple iPhone True Depth IR system.
This allows depth measurements to be taken and the state of human health to 
be determined~\cite{art:deep_learning_based_cross_spectral}.

In the field of distributed multi-modal camera systems, a huge challenge can be
the different spatial resolution of the various cameras.
In contrast to recently published
methods like the Camera Array for Multi-Spectral
Imaging~\cite{art:CAMSI}, which contains identical cameras, we
combine an \ac{IR} camera with a color
camera in our setup.
The combination of
\ac{IR} camera and color camera leads to the problem of different spatial
resolutions that may differ by orders of magnitude.
In practice, the \ac{IR} cameras have lower
spatial resolution as an identically constructed color camera.
A huge task in such a setup is to increase the spatial resolution of an
\ac{IR} image to the spatial resolution of a color image.
In order to increase spatial resolution there are different approaches,
which will be discussed next.

On the one hand, the field of \ac{SISR} is well known.\ \ac{SISR}
reconstructs a \acl{HR} image from a \acl{LR} image~\cite{art:SISR}.\ Solving
the problem of \ac{SISR} in research is very common and there are
two different concepts.
First, different interpolation methods such as bicubic interpolation
can be mentioned among many others.
Bicubic interpolation is a
common method in the field of \ac{SISR} which is well
studied and introduces only minor computational
complexity~\cite{art:sum_inter}.
Moreover, deep neuronal networks can be mentioned.
\ac{VDSR} was one important milestone in the field of
\ac{SISR}.\ One important aspect is, that \ac{VDSR} uses an interpolated
image as input.\ The main idea to train \ac{VDSR} is to learn the residual
between the output image and the input \mbox{image~\cite{net:VDSR, art:deep_spatial_interpolation_rain_field}}.
Furthermore, \ac{EDSR} should be mentioned.\ In
contrast to \ac{VDSR}, \ac{EDSR} does not use an interpolated image, instead
the network uses the \acl{LR} image directly.\ In order to use the image
directly, \ac{EDSR}
was trained separately for each scaling~\cite{net:EDSR,art:deep_learning_based_sisr_review}.

On the other hand, the field of \ac{MISR} can be
named.\ To
achieve a higher spatial resolution of an image, \ac{MISR} combines multiple images from the
same scene~\cite{art:SR_diff_exposed_MR}.\ To capture multiple images from one
scene,
there are various
options.\ One is \ac{VSR}.\ In \ac{VSR}, the
datasets are ordered frames of the same scene.\ To achieve a \acl{HR}
image at time $t=T$, temporally adjacent images can be used~\cite{art:DVSR}.
Another approach is \ac{SSR}.\ \ac{SSR} captures two images from different
positions simultaneously from the same scene.\ Among other things, this leads
to the parallax effect.\ This effect has to be
compensated by \mbox{algorithms~\cite{art:SSR_parallax, art:deep_learning_for_SR}}.

Since capturing \acl{IR} images requires larger exposure times, one can
receive huge movements between consecutive images.\ This
increases the difficulty to find corresponding image regions between
consecutive
images~\cite{art:DVSR}.\ Capturing a moving scene with large exposure times
results in motion blur~\cite{art:handling_motion_blur}.\ Because of that,
\ac{VSR} is not an option in our considered scenario.\ That is why we go into
the field of \ac{SSR}.\
Unlike~\cite{art:hybrid_sisr_misr} and~\cite{art:a_robust_stereo_matching}, we have
a \acl{HR} color image and a \acl{LR} \ac{IR} image.

\section{\acf{GIRRE}}\label{sec:prop_meth}
In the following, we introduce our \acf{GIRRE} method.
The algorithm increases the spatial resolution of the \acl{LR} \ac{IR}
image $\widetilde{X}$ to the same spatial resolution as that of the \acl{HR}
color image $G$.
We obtain the approximated image $\widecheck{X}$.
Similar to~\cite{art:guided_filter}, we assume
that the enhanced image $\widehat{X}$ is a local linear transformation of
the \acl{HR} color image $G$.
Moreover, \ac{GIRRE} uses the guided filter that was introduced
in~\cite{art:guided_filter} to transfer the information contained in the
\acl{HR} color image $G$ to the approximated image $\widecheck{X}$.
In the following, we describe the proposed \ac{GIRRE} algorithm.

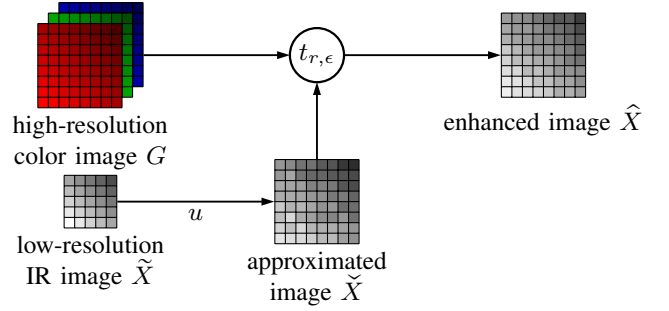
\begin{figure}[t]
\begin{subfigure}[]{.32\columnwidth}
\centering
\begin{tikzpicture}[y=1cm, x=1cm, yscale=\globalscale,
	xscale=\globalscale, inner sep=0pt, outer sep=pt, remember picture]

\foreach \x in {0,1,...,\sizex}{
	\foreach \y in {0,1,...,\sizey}{
			\pgfmathparse{100 - (\x+\y) / (\sizex+\sizey) * 70}
			\fill[blue!\pgfmathresult!black] (\x+2,\y+2) rectangle ++(1,1);
		}
	}
\draw[step=1, thin] (2,2) grid (\sizex+3,\sizey+3);

\foreach \x in {0,1,...,\sizex}{
	\foreach \y in {0,1,...,\sizey}{
			\pgfmathparse{100 - (\x+\y) / (\sizex+\sizey) * 70}
			\fill[green!\pgfmathresult!black] (\x+1,\y+1) rectangle ++(1,1);

		}
	}

\draw[step=1, thin] (1,1) grid (\sizex+2,\sizey+2);

\foreach \x in {0,1,...,\sizex}{
	\foreach \y in {0,1,...,\sizey}{
			\pgfmathparse{100 - (\x+\y) / (\sizex+\sizey) * 70}
			\fill[red!\pgfmathresult!black] (\x,\y) rectangle ++(1,1);
		}
	}

\draw[step=1, thin] (0,0) grid (\sizex+1,\sizey+1);

\coordinate (rgb_e) at (current bounding box.east);
\coordinate (rgb_n) at (current bounding box.north);
\coordinate (rgb_w) at (current bounding box.west);
\coordinate (rgb_s) at (current bounding box.south);

\end{tikzpicture}
\end{subfigure}
\hfill
\begin{subfigure}[]{.32\columnwidth}
\centering

\begin{tikzpicture}[y=1cm, x=1cm, yscale=\globalscale, xscale=\globalscale, inner sep=0pt, outer sep=0pt, remember picture]

\node[dspmultiplier, scale=1.8] (gf) {};

\coordinate (gf_e) at (current bounding box.east);
\coordinate (gf_n) at (current bounding box.north);
\coordinate (gf_w) at (current bounding box.west);
\coordinate (gf_s) at (current bounding box.south);

\end{tikzpicture}
\end{subfigure}
\hfill
\begin{subfigure}[]{.32\columnwidth}
\centering

\begin{tikzpicture}[y=1cm, x=1cm, yscale=\globalscale, xscale=\globalscale, inner sep=0pt, outer sep=0pt, remember picture]

\foreach \x in {0,1,...,\sizex}{
	\foreach \y in {0,1,...,\sizey}{
			\pgfmathparse{100 - (((\x+\y) / (\sizex+\sizey) * 70) + 7 * rnd)}
			\fill[white!\pgfmathresult!black] (\x,\y) rectangle ++(1,1);

		}
	}

\draw[step=1, thin] (0,0) grid (\sizex+1,\sizey+1);

\coordinate (gfi_e) at (current bounding box.east);
\coordinate (gfi_n) at (current bounding box.north);
\coordinate (gfi_w) at (current bounding box.west);
\coordinate (gfi_s) at (current bounding box.south);

\end{tikzpicture}
\end{subfigure}

\vspace{2em}

\begin{subfigure}[]{.32\columnwidth}
\center
\begin{tikzpicture}[y=1cm, x=1cm, yscale=\globalscale, xscale=\globalscale, inner sep=0pt, remember picture]

\foreach \x in {0,1,...,\sizexs}{
	\foreach \y in {0,1,...,\sizeys}{
			\pgfmathparse{100 - (\x+\y) / (\sizexs+\sizeys) * 70}
			\fill[white!\pgfmathresult!black] (\x,\y) rectangle ++(1,1);
		}
	}

\draw[step=1, thin] (0,0) grid (\sizexs+1,\sizeys+1);

\coordinate (source_e) at (current bounding box.east);
\coordinate (source_n) at (current bounding box.north);
\coordinate (source_w) at (current bounding box.west);
\coordinate (source_s) at (current bounding box.south);

\end{tikzpicture}
\end{subfigure}
\hfill
\begin{subfigure}[]{.32\columnwidth}
\centering
\begin{tikzpicture}[y=1cm, x=1cm, yscale=\globalscale, xscale=\globalscale, inner sep=0pt, outer sep=0pt, remember picture]

\foreach \x in {0,1,...,\sizex}{
	\foreach \y in {0,1,...,\sizey}{
			\pgfmathparse{100 - (((\x+\y) / (\sizex+\sizey) * 70) + 15 * rnd)}
			\fill[white!\pgfmathresult!black] (\x,\y) rectangle ++(1,1);

		}
	}

\draw[step=1, thin] (0,0) grid (\sizex+1,\sizey+1);

\coordinate (up_e) at (current bounding box.east);
\coordinate (up_n) at (current bounding box.north);
\coordinate (up_w) at (current bounding box.west);
\coordinate (up_s) at (current bounding box.south);

\end{tikzpicture}
\end{subfigure}
\hfill
\begin{subfigure}[]{.32\columnwidth}
\centering
\begin{tikzpicture}[y=1cm, x=1cm, yscale=\globalscale, xscale=\globalscale, inner sep=0pt, outer sep=0pt, remember picture]
\end{tikzpicture}
\end{subfigure}

\begin{tikzpicture}[overlay, remember picture,y=1cm, x=1cm, yscale=\globalscale,
	xscale=\globalscale, inner sep=0pt, outer sep=0pt,
	every text node part/.style={align=center}]
	
	\draw [dspconn] (source_e) -- (up_w) node [pos=.5, below = 1mm] {$u$};
	\draw [dspconn] (up_n) -- (gf_s);
	\draw [dspconn] (rgb_e) -- (gf_w);
	\draw [dspconn] (gf_e) -- (gfi_w);

	\node [below = 1mm of source_s] {low-resolution\\ IR image $\widetilde{X
	}$};
	\node [below = 1mm of up_s] {approximated\\ image $\widecheck{X}$};
	\node [below = 1mm of rgb_s] {high-resolution\\ color
	image $G$};
	\node at (gf.center) {$t_{r,\epsilon}$};
	\node [below = 1mm of gfi_s] {enhanced image $\widehat{X}$};
\end{tikzpicture}
\vspace{3em}
\caption{First, the proposed method \acs{GIRRE} increases the spatial resolution of
	$\widetilde{X}$.\ After that, the function
	$t_{r,\epsilon}$ uses the color image $G$ to generate details in the
	approximated IR	image $\widecheck{X}$.}
\label{fig:pipeline}
\end{figure}

Because the method is independent of the color depth, we assume the
image values to be between zero and one and are of floating point accuracy.
As shown in
\Cref{fig:pipeline}, a function $u$ upscales the \acl{LR} \ac{IR}
image $\widetilde{X}$ to the size of the \acl{HR} color image $G$
to obtain the approximated image $\widecheck{X}$:
\begin{IEEEeqnarray}{c}
    u(\widetilde{X}) = \widecheck{X}. \label{eq:app_image}
\end{IEEEeqnarray}
Afterwards, we use a transfer function $t_{r,\epsilon}$ to increase the
image
quality of the approximated image $\widecheck{X}$ with the help of the \acl{HR}
color image $G$ to receive the enhanced image $\widehat{X}$:
\begin{IEEEeqnarray}{c}
    t_{r,\epsilon}(\widecheck{X},G) = \widehat{X}. \label{eq:trans}
\end{IEEEeqnarray}
The transfer function~$t_{r,\epsilon}$ accepts two parameters.\ The
radius~$r$, which describes a square window~$\omega_k$ centered around the
considered pixel $k$ and $\epsilon$ as regularization parameter.
Next, we assume the enhanced image $\widehat{X}$ is a local linear
transformation of the guide image $G$:
\begin{IEEEeqnarray}{c}
    \widehat{X}_i = a_kG_i + b_k, \forall i\in\omega_k, \label{eq:loc_lin}
\end{IEEEeqnarray}
where $(a_k,b_k)$ are linear coefficients which are constant in the window $\omega_k$
and $i$ being the pixels in the window.
The local linear model ensures that $\widehat{X}$ has an
edge if $G$ has an edge, because $\nabla\widehat{X}=a\nabla G$.
To
calculate the coefficients, we need the following assumption.\ The enhanced
image $\widehat{X}_i$ is equal to the approximated image $\widecheck{X}_i$
subtracting some distortions $\eta_i$ like noise:
\begin{IEEEeqnarray}{lCl}
     \widehat{X}_i & = \widecheck{X}_i - \eta_i.
\end{IEEEeqnarray}
To achieve minimal distortions, we minimize the difference between
$\widehat{X}$
and $\widecheck{X}$ while keeping the linear model~\eqref{eq:loc_lin}.\
Especially, in the window $\omega_k$, we minimize the cost function:
\begin{IEEEeqnarray}{c}
    E(a_k,b_k) = \sum_{i\in\omega_k}\left(\left(a_kG_i + b_k -
    \widecheck{X}_i\right)^2
    + \epsilon a_k^2\right). \label{eq:lin_ridge}
\end{IEEEeqnarray}
The parameter $\epsilon$ controls the influence of large $a_k$.\ Equation~\eqref{eq:lin_ridge}
is the \textit{linear ridge regression}
model~\cite{book:elements_of_statistical_learning,book:applied_regression_analysis}
with the solution:\
\begin{IEEEeqnarray}{c}
    a_k = \frac{\frac{1}{|\omega_k|}\sum_{i\in\omega_k}G_i \widecheck{X}_i -\mu_k
    \bar{X}_k}{\sigma^2_k + \epsilon},\label{eq:ak}
\end{IEEEeqnarray}
\begin{IEEEeqnarray}{c}
    b_k = \bar{X}_k - a_k\mu_k,\label{eq:bk}
\end{IEEEeqnarray}
with $|\omega_k|$ being the cardinality of $\omega_k$ and $\mu_k$, $\sigma^2_k$ the
mean
and variance of $G$ in $\omega_k$ and $\bar{X}_k$ the mean of $\widecheck{X}$
in $\omega_k$.\ After calculating the parameters $a_k$ and $b_k$, we are
able to compute the
output $\widehat{X}_i$ with respect to~\eqref{eq:loc_lin}.

The problem is, that a pixel~$k$ is involved in all overlapping windows $\omega_i$
which contain~$k$.
Here, $\omega_i$ denote a window at a random pixel $i$.
Thus, we receive different values for $\widehat{X}_k$
in~\eqref{eq:loc_lin} for different windows.
Consequently, the strategy is to average all
the values of $\widehat{X}_k$:
\begin{IEEEeqnarray}{c}
    \widehat{X}_k = \frac{1}{|\omega_i|} \sum_{i|k\in\omega_i}(a_iG_k + b_i).
    \label{eq:avg_yi}
\end{IEEEeqnarray}
Because of the symmetrical window $\omega_i$, the following applies:
\begin{IEEEeqnarray}{c}
    \sum_{i|k\in{\omega_i}} a_i = \sum_{i\in\omega_k} a_i.
    \label{eq:symmetric_window}
\end{IEEEeqnarray}
With~\eqref{eq:symmetric_window}, we can write~\eqref{eq:avg_yi} as:
\begin{IEEEeqnarray}{c}
    \widehat{X}_k = \bar{a_k} G_k + \bar{b_k}, \label{eq:avg_Xi}
\end{IEEEeqnarray}
with $\bar{a}_k=\frac{1}{|\omega_k|}\sum_{i\in\omega_k} a_i$ and
$\bar{b}_k=\frac{1}{|\omega_k|}\sum_{i\in\omega_k} b_i$
being the average coefficients of all windows which overlap the
pixel~$k$.
A pseudocode for \ac{GIRRE} is shown in~\Cref{alg:pseudo}.
\begin{algorithm}[t]
    \caption{Proposed GIRRE algorithm}
    \begin{algorithmic}
        \Input
            \Desc{$\widetilde{X}$}{\acl{LR} \acl{IR} image}
            \Desc{$G$}{\acl{HR} color image}
        \EndInput
        \Output
            \Desc{$\widehat{X}$}{enhanced image}
        \EndOutput
    \end{algorithmic}
    \begin{algorithmic}[1]
        \State upscale $\widetilde{X}$ to the size of $G$ in~\eqref{eq:app_image}
        \ForAll{$i \in |\widecheck{X}|$}\Comment($|\widecheck{X}|$:
        cardinality of $\widecheck{X}$)
            \State compute $a_i$ in~\eqref{eq:ak}
            \State compute $b_i$ in~\eqref{eq:bk}
        \EndFor
        \For{$k \in |\widecheck{X}|$}
            \State compute the average of the $a_i$, such that $i\in\omega_k$
            \State compute the average of the $b_i$, such that $i\in\omega_k$
            \State compute $\widehat{X}_k$ in~\eqref{eq:avg_Xi}
        \EndFor
    \end{algorithmic}
    \label{alg:pseudo}
\end{algorithm}

To sum up, \ac{GIRRE} scales up the spatial resolution of a \acl{LR} \acl{IR} image with the help of a \acl{HR}
color image.
Therefore, we assume a local linear dependency of the images in order to be
able to apply the presented algorithm.

\section{Evaluation}\label{sec:evaluation}
Before evaluating the proposed \ac{GIRRE} method, we will discuss the employed
parameters.
In literature, it is common to use scaling factors of $2\times 2$,
$3\times 3$ and $4\times 4$~\cite{art:comb_si_and_mi_sr,art:PIRM2018}.\
Moreover we use a scaling of $8\times 8$, because color images contain up
to 100 times more pixels than \acl{IR} images so that our calculations
become even more practical.
A scaling $x\times y$ means,
that we scale the width by a factor of $x$ and the height by a factor of $y$.\

For evaluation, we selected the databases CAVE~\cite{ds:CAVE},
Harvard~\cite{ds:HARVARD} and TokyoTech~\cite{ds:TOKYOTECH}.
The datasets are chosen, because the different spectra are captured with a
tunable filter.\ For this reason, we can assume the images
to be well registered.\ The CAVE dataset contains grayscale images between
400\,nm and 700\,nm and a color image which is rendered by the grayscale
images at different wavelengths.\ Furthermore, the Harvard dataset contains
grayscale images between
420\,nm and 720\,nm but without color images.\ TokyoTech contains images
in the range of 420\,nm to 650\,nm~(visible) and for 650\,nm to
1000\,nm~(\acs{IR}).
Like Harvard, TokyoTech does not include a color image.
For the TokyoTech and Harvard datasets, we use~\cite{met:hsi2rgb} to generate
color images.
The image size of the CAVE and TokyoTech datasets are $512\times512$ pixels.
The images of the Harvard dataset are resized to the same spatial resolution.
Furthermore, for visual examples we use a spatial resolution of $512\times512$ pixels, too.
To simulate the
\acl{LR} \ac{IR} image $\widetilde{X}$, we downscale the respective \ac{IR} ground truth
image by the considered scale factor.
For the analysis, we select the image with the longest wavelength from each of the data sets as the
ground truth image.

As shown in~\Cref{sec:prop_meth}, radius~$r$ and the regularization
parameter~$\epsilon$ are essential for
the transfer function $t_{r,\epsilon}$.\ Irrespective of the scale factor, the
evaluation with the TokyoTech database has shown that we obtain the
best results for $\epsilon = 0.1^4$.
The radius $r$ behaves differently from $\epsilon$ since it
depends on the scale factor and the upscale function $u$.
In order to find the radius, we therefore choose TokyoTech as reference
dataset for all functions $u$.\ The
chosen radius has the biggest difference in terms of \ac{PSNR} between the
approximated
image and \ac{GIRRE}.\ The determined parameters are shown
in~\Cref{tab:par}.

\begin{table}[t]
    \centering
    \caption{Parameters for the transfer function\,$t_{r,\epsilon}$ of the proposed GIRRE method
            optimized on TokyoTech.
        Parameter $\epsilon$
        has a fixed value of $0.1^4$. The value for the radius depends on
        the scaling and the procedure how the approximated
        image\,$\widecheck{X}$ is computed. The table shows the radius per
        scaling for $\widecheck{X}_\mathrm{BICUBIC}$ and
        $\widecheck{X}_\mathrm{VDSR}$.}
    \label{tab:par}
    \begin{subtable}[h]{0.3\columnwidth}
        \centering
        \begin{tabular}{c | c}
            \textbf{scale} & \textbf{radius} \\
            \hline
            $2\times 2$ & 1 \\
            $3\times 3$ & 3 \\
            $4\times 4$ & 6 \\
            $8\times 8$ & 15
        \end{tabular}
        \vspace*{1mm}
        \caption{$\widecheck{X}_\mathrm{BICUBIC}$}
        \label{par:bicubic}
    \end{subtable}
    \begin{subtable}[h]{0.3\columnwidth}
        \centering
        \begin{tabular}{c | c}
            \textbf{scale} & \textbf{radius} \\
            \hline
            $2\times 2$ & 2 \\
            $3\times 3$ & 3 \\
            $4\times 4$ & 4 \\
            $8\times 8$ & 15
        \end{tabular}
        \vspace*{1mm}
        \caption{$\widecheck{X}_\mathrm{VDSR}$}
        \label{par:vdsr}
    \end{subtable}
\end{table}

The following evaluations are carried out on the Harvard and CAVE datasets, which
are independent of the TokyoTech dataset that has been used for parameter
optimization.
\Cref{tab:bicubic} shows the \ac{PSNR} in \ac{db} for our proposed
\ac{GIRRE} method with bicubic interpolation as upscale function~$u$ and for
bicubic interpolation as competing method.
The last column shows the
difference between \ac{GIRRE} and bicubic interpolation.\ If the
difference is positive, our method increases in terms of the \ac{PSNR}.\ When
it is negative, our method decreases the \ac{PSNR}.\ For the table, we have a
multi
index.\ This means, that we group the table by scale factor, radius and
dataset.\ The radius has a one-to-one relationship with the scale factor.\
As last level, we have different datasets and an extra row for the
average value in each group.\ We highlight the best method in bold letters.
Furthermore, the radius increases by enlarging the scaling.\ This can
be explained by the fact that the function~$t_{r,\epsilon}$ requires more
information to enhance the image.
From~\Cref{tab:bicubic}, we can infer that
the proposed method significantly increases \ac{PSNR} for all datasets
by scaling $2\times 2$ up to $8\times 8$.
The gain in terms of the \ac{PSNR} is between 0.57\,\acs{db} and 2.28\,\acs{db}.
For bicubic interpolation as upscale
function $u$, we increase the \ac{PSNR} on average by 1.39\,\acs{db}.
\begin{table}[t]
\centering
\caption{Comparison of the proposed GIRRE algorithm with respect to BICUBIC\hspace{1sp}\cite{art:sum_inter}
on multiple datasets in terms of PSNR.}
\label{tab:bicubic}
\begin{tabular}{|c|c|c||ccc|}
 \hline
\bfseries Scale \bfseries & \bfseries Radius & \bfseries Dataset & \bfseries
 GIRRE & \bfseries BICUBIC\hspace{1sp}\cite{art:sum_inter} & \bfseries DIFF \\
 \hline \hline
\multirow[c]{3}{*}{2 $\times$ 2} & \multirow[c]{3}{*}{1} & Harvard\hspace{1sp}\cite{ds:HARVARD} &
 \bfseries 34.22 & 33.63 & \bfseries 0.59 \\
 &  & CAVE\hspace{1sp}\cite{ds:CAVE} & \bfseries 41.19 & 40.62 & \bfseries 0.57 \\
 &  & Average & \bfseries 37.71 & 37.12 & \bfseries 0.59 \\
 \hline
\multirow[c]{3}{*}{3 $\times$ 3} & \multirow[c]{3}{*}{3} & Harvard\hspace{1sp}\cite{ds:HARVARD} &
 \bfseries 32.47 & 31.25 & \bfseries 1.22 \\
 &  & CAVE\hspace{1sp}\cite{ds:CAVE} & \bfseries 38.12 & 36.74 & \bfseries 1.38 \\
 &  & Average & \bfseries 35.30 & 34.00 & \bfseries 1.30 \\
 \hline
\multirow[c]{3}{*}{4 $\times$ 4} & \multirow[c]{3}{*}{6} & Harvard\hspace{1sp}\cite{ds:HARVARD} &
 \bfseries 31.45 & 29.93 & \bfseries 1.52 \\
 &  & CAVE\hspace{1sp}\cite{ds:CAVE} & \bfseries 36.44 & 34.56 & \bfseries 1.88 \\
 &  & Average & \bfseries 33.95 & 32.25 & \bfseries 1.70 \\
 \hline
\multirow[c]{3}{*}{8 $\times$ 8} & \multirow[c]{3}{*}{15} & Harvard\hspace{1sp}\cite{ds:HARVARD} &
 \bfseries 29.08 & 27.38 & \bfseries 1.70 \\
 &  & CAVE\hspace{1sp}\cite{ds:CAVE} & \bfseries 32.14 & 29.86 & \bfseries 2.28 \\
 &  & Average & \bfseries 30.61 & 28.62 & \bfseries 1.99 \\
 \hline
\end{tabular}
\end{table}

In the following, we compare our novel \ac{GIRRE} method to VDSR\nobreak\hspace{1sp}\cite{net:VDSR}, which
is a recent super-resolution method based on a neuronal network.
\Cref{tab:vdsr} has the same structure as \Cref{tab:bicubic}.
Here, the approximated image $\widecheck{X}$ was generated with the upscale
function $u = \text{VDSR\nobreak\hspace{1sp}\cite{net:VDSR}}$.
In this case, we increase \ac{PSNR} by at least 1.75\,\acs{db}.
For scaling $2 \times 2$ and the Harvard dataset, we achieve the best
result with a difference of 2.79\,\acs{db}.
On average over all datasets, we increase \ac{PSNR} by 2.37\,\acs{db}
with \ac{GIRRE} compared to VDSR\nobreak\hspace{1sp}\cite{net:VDSR}.

\begin{table}[t]
\centering
\caption{Comparison of the proposed GIRRE algorithm with respect to VDSR\hspace{1sp}\cite{net:VDSR}
on multiple datasets in terms of PSNR.}
\label{tab:vdsr}
\begin{tabular}{|c|c|c||ccc|}
\hline
\bfseries Scale & \bfseries Radius & \bfseries Dataset & \bfseries GIRRE & \bfseries VDSR\hspace{1sp}\cite{net:VDSR} & \bfseries DIFF \\
\hline
\hline
\multirow[c]{3}{*}{2 $\times$ 2} & \multirow[c]{3}{*}{2} & Harvard\hspace{1sp}\cite{ds:HARVARD} & \bfseries 36.23 & 33.44 & \bfseries 2.79 \\
 &  & CAVE\hspace{1sp}\cite{ds:CAVE} & \bfseries 41.83 & 39.55 & \bfseries 2.28 \\
 &  & Average & \bfseries 39.03 & 36.49 & \bfseries 2.54 \\
\hline
\multirow[c]{3}{*}{3 $\times$ 3} & \multirow[c]{3}{*}{3} & Harvard\hspace{1sp}\cite{ds:HARVARD} & \bfseries 33.86 & 31.13 & \bfseries 2.73 \\
 &  & CAVE\hspace{1sp}\cite{ds:CAVE} & \bfseries 39.24 & 36.63 & \bfseries 2.61 \\
 &  & Average & \bfseries 36.55 &  33.88 & \bfseries 2.67 \\
\hline
\multirow[c]{3}{*}{4 $\times$ 4} & \multirow[c]{3}{*}{4} & Harvard\hspace{1sp}\cite{ds:HARVARD} & \bfseries 32.34 & 30.33 & \bfseries 2.01 \\
 &  & CAVE\hspace{1sp}\cite{ds:CAVE} & \bfseries 37.60 & 35.23 & \bfseries 2.37 \\
 &  & Average & \bfseries 34.97 & 32.78 & \bfseries 2.19 \\
\hline
\multirow[c]{3}{*}{8 $\times$ 8} & \multirow[c]{3}{*}{15} & Harvard\hspace{1sp}\cite{ds:HARVARD} & \bfseries 29.11 & 27.36 & \bfseries 1.75 \\
 &  & CAVE\hspace{1sp}\cite{ds:CAVE} & \bfseries 32.18 & 29.77 & \bfseries 2.41 \\
 &  & Average & \bfseries 30.64 & 28.57 & \bfseries 2.07 \\
 \hline
\end{tabular}
\end{table}

\Cref{tab:bicubic,tab:vdsr} show, that on all considered scalings the
proposed \ac{GIRRE} method with upscale function $u = \text{VDSR}$ achieves
a higher \ac{PSNR} in contrast to \ac{GIRRE} with upscale function
$u = \text{BICUBIC}$.
That is why we choose $u = \text{VDSR}$ for further
evaluations.

\Cref{tab:scaling} shows the \ac{PSNR} in \ac{db} and the \ac{SSIM} in
brackets of
\ac{GIRRE} and the \mbox{state-of-the-art} method
BICUBIC\nobreak\hspace{1sp}\cite{art:sum_inter}.
Furthermore, since BICUBIC\nobreak\hspace{1sp}\cite{art:sum_inter} is a simple and old
method, we also compare \ac{GIRRE} with \ac{EDSR}\nobreak\hspace{1sp}\cite{net:EDSR}.
The difference in \ac{PSNR} increases from scaling $2\times 2$ to $4\times 4$
and then decreases.
Furthermore, the difference of the \ac{SSIM} increases from the smallest
considered scaling to the largest considered scaling.
This behavior is observed for both \mbox{state-of-the-art} methods.
Furthermore, \ac{GIRRE} has the best \ac{PSNR} and \ac{SSIM} over the
\mbox{state-of-the-art} methods for scaling $2\times 2$ to $8\times 8$ which we mark as bold.

\begin{table}[t]
\centering
\caption{Comparison of the proposed GIRRE method with bicubic
interpolation\hspace{1sp}\cite{art:sum_inter} and EDSR\hspace{1sp}\cite{net:EDSR}.
The table shows the PSNR\,(SSIM) scores.
}
\label{tab:scaling}
\begin{tabular}{|c||ccc|c}
 \hline
 \bfseries Scale \bfseries & \bfseries GIRRE & \bfseries BICUBIC\hspace{1sp}\cite{art:sum_inter} & \bfseries EDSR\hspace{1sp}\cite{net:EDSR} \\
 \hline
 \hline
 2 $\times$ 2 & \bfseries 39.03\,(0.95) & 37.12\,(0.93) & 38.47\,(0.94) \\
 3 $\times$ 3 & \bfseries 36.55\,(0.92) & 34.00\,(0.88) & 35.18\,(0.89) \\
 4 $\times$ 4 & \bfseries 34.97\,(0.90) & 32.25\,(0.84) & 33.20\,(0.85) \\
 8 $\times$ 8 & \bfseries 30.64\,(0.86) & 28.62\,(0.76) & 29.65\,(0.78) \\
 \hline
\end{tabular}
\end{table}

To conclude the evaluation, we give some visual examples.
\Cref{fig:chart_and_stuffed} \input{examples_cave_chart_and_stuffed.tikz}
shows a section of the \textit{chart\_and\_stuffed\_toy} image of the CAVE dataset.
The first row shows the ground truth image, the low-resolution \ac{IR} image and the enhanced image \ac{GIRRE}.
The high-resolution color image, the enhanced image BICUBIC\nobreak\hspace{1sp}\cite{art:sum_inter} and the enhanced
image \ac{EDSR}\nobreak\hspace{1sp}\cite{net:EDSR}
are shown at the second row.
Furthermore, we added the \ac{PSNR} scores to the images and
mark the highest \ac{PSNR} value in bold letters.
To achieve the processed images, we use a $4\times 4$ scaling.
This results in a spatial resolution of $128\times128$ of the low-resolution \ac{IR} image.
The image section shows the hair of a stuffed toy as well as a chart.
For this example, our method achieves
33.55\,dB\@.
The best \mbox{state-of-the-art} method
\ac{EDSR}\nobreak\hspace{1sp}\cite{net:EDSR} achieves
only 31.00\,dB, which leads to a difference of 2.55\,dB.\ This can
be noticed for all \mbox{state-of-the-art} methods.\ For example, the
numbers 3, 4, and 5 can only be recognized on the image
reconstructed by \ac{GIRRE}.

Finally, \Cref{fig:harvard_512} \input{examples_harvard.tikz}
shows \textit{img1} of the Harvard dataset.
The structure is like in~\Cref{fig:chart_and_stuffed}, but using an $8\times 8$
scaling.
The \mbox{low-resolution} IR image has a resolution of $64\times64$.
In the middle of the image, we can see a ventilation inlet.
Our method shows sharp edges for the
ventilation inlet.\ In this example, we can increase \ac{PSNR} by 2.32
\ac{db} with respect to the best \mbox{state-of-the-art} method \ac{EDSR}\nobreak\hspace{1sp}\cite{net:EDSR}.

\section{Conclusion}\label{sec:conclusion}
Modern electronic devices such as smartphones use multi-modal stereo cameras.
Especially, the combination of a color camera and an \ac{IR} camera is
widely employed.
However, a major problem is the huge difference between the spatial
resolution of current sensors.
Recent color sensors typically take 100 times more pixels than current
\acl{IR} sensors.
For this reason, color images contain significantly more spatial information
of the scene than infrared images.

To this end, we introduce the novel \acf{GIRRE} algorithm to increase the
quality of \acl{LR}
\acl{IR} images with the help of \acl{HR} color images.
In~\Cref{tab:scaling}, we showed that our method achieves an average gain of
1.2\,\acs{db} and
at best we can increase the \acs{PSNR} by 1.8\,\acs{db} compared to \ac{EDSR}\nobreak\hspace{1sp}\cite{net:EDSR}.
Furthermore, we demonstrated the superiority of our method by giving visual
examples.

In sum, the quality of \acl{LR} \acl{IR}
images can be significantly improved with the aid of \acl{HR} color images
when using the proposed \acs{GIRRE} algorithm.

\newpage

\balance
\bibliographystyle{IEEEtran}
\bibliography{IEEEabrv,IEEE}

\begin{thebibliography}{10}
\providecommand{\url}[1]{#1}
\csname url@samestyle\endcsname
\providecommand{\newblock}{\relax}
\providecommand{\bibinfo}[2]{#2}
\providecommand{\BIBentrySTDinterwordspacing}{\spaceskip=0pt\relax}
\providecommand{\BIBentryALTinterwordstretchfactor}{4}
\providecommand{\BIBentryALTinterwordspacing}{\spaceskip=\fontdimen2\font plus
\BIBentryALTinterwordstretchfactor\fontdimen3\font minus
  \fontdimen4\font\relax}
\providecommand{\BIBforeignlanguage}[2]{{%
\expandafter\ifx\csname l@#1\endcsname\relax
\typeout{** WARNING: IEEEtran.bst: No hyphenation pattern has been}%
\typeout{** loaded for the language `#1'. Using the pattern for}%
\typeout{** the default language instead.}%
\else
\language=\csname l@#1\endcsname
\fi
#2}}
\providecommand{\BIBdecl}{\relax}
\BIBdecl

\bibitem{on:IMX990}
\BIBentryALTinterwordspacing
{SWIR} {I}mage {S}ensor. Accessed: 23.03.2023. [Online]. Available:
  \url{https://www.sony-semicon.com/en/products/is/industry/swir.html}
\BIBentrySTDinterwordspacing

\bibitem{on:IMX411}
\BIBentryALTinterwordspacing
{R}olling {S}hutter {I}mage {S}ensor. [Online]. Available:
  \url{https://www.sony-semicon.com/en/products/is/industry/rolling-shutter.html}
\BIBentrySTDinterwordspacing

\bibitem{art:CAMSI}
N.~Genser, J.~Seiler, and A.~Kaup, ``Camera array for multi-spectral imaging,''
  \emph{IEEE Transactions on Image Processing}, vol.~29, pp. 9234--9249, 2020.

\bibitem{art:deep_material_aware}
T.~Zhi, B.~R. Pires, M.~Hebert, and S.~G. Narasimhan, ``Deep material-aware
  cross-spectral stereo matching,'' in \emph{Proc. IEEE/CVF Conference on
  Computer Vision and Pattern Recognition (CVPR)}, 2018, pp. 1916--1925.

\bibitem{pat:camera_module_including}
Y.~Hwang, K.~Byon, and J.~Kim, ``Camera module including multi-lens and
  electronic device having the same,'' Patent WO/2017/048\,081, Mar. 2017.

\bibitem{pat:dual_camera_module}
W.~K. Park, J.~W. Kim, J.~Y. Park, S.~J. Hwang, I.~C. Shim, D.~H. Jeong, B.~G.
  An, D.~W. Kim, and K.~H. Choi, ``Dual camera module and portable electronic
  device,'' U.S. Patent 11,048,307, Jun. 2021.

\bibitem{art:deep_learning_based_cross_spectral}
N.~Genser, A.~Spruck, J.~Seiler, and A.~Kaup, ``Deep learning based
  cross-spectral disparity estimation for stereo imaging,'' in \emph{Proc. IEEE
  International Conference on Image Processing (ICIP)}, 2020, pp. 2536--2540.

\bibitem{art:SISR}
M.~Irani and S.~Peleg, ``Improving resolution by image registration,''
  \emph{CVGIP: Graphical Models and Image Processing}, vol.~53, no.~3, pp.
  231--239, 1991.

\bibitem{art:sum_inter}
I.~Amidror, ``{Scattered data interpolation methods for electronic imaging
  systems: a survey},'' \emph{Journal of Electronic Imaging}, vol.~11, no.~2,
  pp. 157 -- 176, 2002.

\bibitem{net:VDSR}
J.~Kim, J.~K. Lee, and K.~M. Lee, ``Accurate image super-resolution using very
  deep convolutional networks,'' in \emph{Proc. IEEE Conference on Computer
  Vision and Pattern Recognition (CVPR)}, 2016, pp. 1646--1654.

\bibitem{art:deep_spatial_interpolation_rain_field}
G.~Yang, Z.~Chen, D.~L. Ndzi, L.~Yang, A.-H. Al-Hassani, D.~C. Paul, Z.~Duan,
  and J.~Chen, ``Deep spatial interpolation of rain field for u.k. satellite
  networks,'' \emph{IEEE Transactions on Antennas and Propagation}, vol.~71,
  no.~2, pp. 1793--1803, 2023.

\bibitem{net:EDSR}
B.~Lim, S.~Son, H.~Kim, S.~Nah, and K.~M. Lee, ``Enhanced deep residual
  networks for single image super-resolution,'' in \emph{Proc. IEEE Conference
  on Computer Vision and Pattern Recognition Workshops (CVPRW)}, 2017, pp.
  1132--1140.

\bibitem{art:deep_learning_based_sisr_review}
K.~Chauhan, S.~N. Patel, M.~Kumhar, J.~Bhatia, S.~Tanwar, I.~E. Davidson, T.~F.
  Mazibuko, and R.~Sharma, ``Deep learning-based single-image super-resolution:
  A comprehensive review,'' \emph{IEEE Access}, vol.~11, pp. 21\,811--21\,830,
  2023.

\bibitem{art:SR_diff_exposed_MR}
T.~Richter and A.~Kaup, ``Super-resolution for differently exposed
  mixed-resolution multi-view images adapted by a histogram matching method,''
  in \emph{Proc. IEEE International Conference on Acoustics, Speech and Signal
  Processing (ICASSP)}, 2017, pp. 2022--2026.

\bibitem{art:DVSR}
X.~Tao, H.~Gao, R.~Liao, J.~Wang, and J.~Jia, ``Detail-revealing deep video
  super-resolution,'' in \emph{Proc. IEEE International Conference on Computer
  Vision (ICCV)}, 2017, pp. 4482--4490.

\bibitem{art:SSR_parallax}
L.~Wang, Y.~Wang, Z.~Liang, Z.~Lin, J.~Yang, W.~An, and Y.~Guo, ``Learning
  parallax attention for stereo image super-resolution,'' in \emph{Proc.
  IEEE/CVF Conference on Computer Vision and Pattern Recognition (CVPR)}, 2019,
  pp. 12\,242--12\,251.

\bibitem{art:deep_learning_for_SR}
Z.~Wang, J.~Chen, and S.~C.~H. Hoi, ``Deep learning for image super-resolution:
  A survey,'' \emph{IEEE Transactions on Pattern Analysis and Machine
  Intelligence}, vol.~43, no.~10, pp. 3365--3387, 2021.

\bibitem{art:handling_motion_blur}
Z.~Ma, R.~Liao, X.~Tao, L.~Xu, J.~Jia, and E.~Wu, ``Handling motion blur in
  multi-frame super-resolution,'' in \emph{2015 IEEE Conference on Computer
  Vision and Pattern Recognition (CVPR)}, 2015, pp. 5224--5232.

\bibitem{art:hybrid_sisr_misr}
M.~Bätz, A.~Eichenseer, J.~Seiler, M.~Jonscher, and A.~Kaup, ``Hybrid
  super-resolution combining example-based single-image and interpolation-based
  multi-image reconstruction approaches,'' in \emph{Proc. IEEE International
  Conference on Image Processing (ICIP)}, 2015, pp. 58--62.

\bibitem{art:a_robust_stereo_matching}
L.~T. Sach, K.~Atsuta, K.~Hamamoto, and S.~Kondo, ``A robust stereo matching
  method for low texture stereo images,'' in \emph{Proc. IEEE-RIVF
  International Conference on Computing and Communication Technologies
  (ICCCT)}, 2009, pp. 1--8.

\bibitem{art:guided_filter}
K.~He, J.~Sun, and X.~Tang, ``Guided image filtering,'' \emph{IEEE Transactions
  on Pattern Analysis and Machine Intelligence}, vol.~35, no.~6, pp.
  1397--1409, 2013.

\bibitem{book:elements_of_statistical_learning}
T.~Hastie, R.~Tibshirani, and J.~Friedman, \emph{The Elements of Statistical
  Learning}.\hskip 1em plus 0.5em minus 0.4em\relax Springer New York, NY,
  2009.

\bibitem{book:applied_regression_analysis}
N.~Draper and H.~Smith, \emph{Applied Regression Analysis, Third
  Edition}.\hskip 1em plus 0.5em minus 0.4em\relax John Wiley \& Sons, Ltd,
  2014.

\bibitem{art:comb_si_and_mi_sr}
T.~Richter, J.~Seiler, W.~Schnurrer, M.~Bätz, and A.~Kaup, ``Combining
  single-image and multiview super-resolution for mixed-resolution image plus
  depth data,'' in \emph{Proc. 23rd European Signal Processing Conference
  (EUSIPCO)}, 2015, pp. 1840--1844.

\bibitem{art:PIRM2018}
M.~Shoeiby, A.~Robles-Kelly, R.~Timofte, R.~Zhou, F.~Lahoud, S.~S{\"u}sstrunk,
  Z.~Xiong, Z.~Shi, C.~Chen, D.~Liu, Z.-J. Zha, F.~Wu, K.~Wei, T.~Zhang,
  L.~Wang, Y.~Fu, K.~Nagasubramanian, A.~K. Singh, A.~Singh, S.~Sarkar, and
  B.~Ganapathysubramanian, ``Pirm2018 challenge on spectral image
  super-resolution: Methods and results,'' in \emph{Proc. Computer Vision --
  ECCV 2018 Workshops}, L.~Leal-Taix{\'e} and S.~Roth, Eds.\hskip 1em plus
  0.5em minus 0.4em\relax Cham: Springer International Publishing, 2019, pp.
  356--371.

\bibitem{ds:CAVE}
F.~Yasuma, T.~Mitsunaga, D.~Iso, and S.~K. Nayar, ``{Generalized Assorted Pixel
  Camera: Postcapture Control of Resolution, Dynamic Range, and Spectrum},''
  \emph{IEEE Transactions on Image Processing}, vol.~19, no.~9, pp. 2241--2253,
  2010.

\bibitem{ds:HARVARD}
A.~Chakrabarti and T.~Zickler, ``{Statistics of Real-World Hyperspectral
  Images},'' in \emph{Proc.~IEEE Conference on Computer Vision and Pattern
  Recognition (CVPR)}, 2011, pp. 193--200.

\bibitem{ds:TOKYOTECH}
Y.~Monno, H.~Teranaka, K.~Yoshizaki, M.~Tanaka, and M.~Okutomi, ``Single-sensor
  rgb-nir imaging: High-quality system design and prototype implementation,''
  \emph{IEEE Sensors Journal}, vol.~19, no.~2, pp. 497--507, 2019.

\bibitem{met:hsi2rgb}
M.~{Magnusson}, J.~{Sigurdsson}, S.~E. {Armansson}, M.~O. {Ulfarsson},
  H.~{Deborah}, and J.~R. {Sveinsson}, ``Creating {RGB} images from
  hyperspectral images using a color matching function,'' in \emph{Proc. IEEE
  International Geoscience and Remote Sensing Symposium (IGARSS)}, 2020, pp.
  2045--2048.

\end{thebibliography}

\end{document}